\documentstyle[aps,floats,twocolumn,prb]{revtex} 
\begin{document} 
\title{ Phase diagram of localization in a magnetic field } 

\author{Thorsten Dr\"ose$^1$, Markus Batsch $^{1,2}$, Isa Kh. Zharekeshev$^1$, 
        and Bernhard Kramer$^1$}

\address{
$^1$ I. Institut f\"ur Theoretische Physik, Universit\"at Hamburg,
Jungiusstrasse~9, D-20355 Hamburg, Germany\\
$^2$ Physikalisch-Technische Bundesanstalt, Bundesallee 100, D-38116
Braunschweig, Germany\smallskip}
\date{\today}
\maketitle
\begin{abstract}
The phase diagram of localization is numerically calculated for a 
three-dimensional disordered system in the presence of a magnetic field 
using the Peierls substitution.
The mobility-edge trajectory shifts in the energy-disorder space when 
increasing the field. In the band center, localized states near the phase 
boundary become delocalized. The obtained field dependence of the critical 
disorder is in agreement with a power-law behavior expected from scaling 
theory.
Close to the tail of the band the magnetic field causes localization
of extended states.
\end{abstract}

\narrowtext

 Though being intensively investigated since the late 1950s \cite{ande58}, 
the problem of the disorder-induced metal-insulator transition (MIT) 
of non-interacting electrons in three dimensions (3D) can be considered as 
being still unsolved. 
There is a controversial discussion, whether the critical behavior at the 
Anderson
transition (AT) can be classified with respect to the fundamental symmetry
of the Hamiltonian as proposed within field theoretical descriptions 
\cite{Wegn89,hika90}.
One would expect that critical properties governed by the exponent of the 
localization length $ \nu$ were altered when the universality class was 
changed.
For instance, by applying an external magnetic field to a system with 
spin rotation invariance, time reversal symmetry is broken and the 
universality class changes from orthogonal to unitary. 
On the basis of considerably reduced error bars recent numerical 
calculations \cite{Slev97} seem to provide evidence that $ \nu$ is
sensitive to symmetry breaking in contrast to previous results 
\cite{kram93,Henn94}.
Therefore, it is particularly interesting to investigate the importance 
of the magnetic field for the localization mechanism that drives the MIT.

In this paper we report results on the influence of a 
homogeneous magnetic field on the {\it phase diagram of localization}
of the disordered Anderson model (AM).
Previous findings concerned the phase diagram 
without magnetic field,
including both exact analytical solutions for the infinite Cayley tree 
(Bethe lattice)~\cite{Abou74,Kawa93}
and numerical studies for a real 3D lattice 
\cite{kram93,bulk85,kroh90a,Grussbach95}. 
Based on numerical results obtained by the transfer-matrix approach
some conclusions on the nature of the 
wave functions and localization properties have been drawn
\cite{kram93,bulk85}:
In the band center, electrons are localized due to quantum interference,
while outside the unperturbed band electrons 
become localized below a certain value of the density of states (DOS)
\cite{bulk85}. Here, we consider the magnetic
field as an additional continuous parameter which can induce the MIT 
\cite{shap84,lee85}.
By using the transfer-matrix method we calculate a complete mobility-edge 
trajectory in the energy-disorder space for a finite magnetic 
field. We find two regimes with 
entirely different behavior with respect to the magnetic-field-driven 
MIT. In the band center, localized states near the zero-field 
phase boundary become extended when applying a magnetic field. In 
slightly disordered metallic systems for states close to the mobility-edge 
the field effect is opposite. 
  
 We use the AM with diagonal disorder 
\cite{ande58} and Peierls phase factors in the hopping matrix elements 
\cite{peie33}, 

\begin{equation}  \label{TBH}
H= \sum_{ {\bf r} } \epsilon_{ {\bf r} }  | {\bf r} \rangle \langle {\bf r} | 
  + \sum_{ {\bf r},{\bf \Delta}  } t_{ {\bf r},{\bf r} + {\bf \Delta}}  
| {\bf r} \rangle \langle {\bf r}+ {\bf \Delta} | .
\end{equation}
Energies are measured in units of the modulus of the hopping matrix elements
$t_{{\bf r}, {\bf r}+{\bf \Delta}}$,
lengths in units of the lattice constant $ a=1 $.
The states $| {\bf r} \rangle $ are associated with the sites 
of a simple cubic lattice. 
The site energies  $ \epsilon_{ {\bf r} }$ are distributed uniformly at 
random between $-W/2 $ and $ W/2 $.
Only hopping between nearest neighbors $ {\bf r} $ and 
$ {\bf r} + {\bf \Delta } $ is taken into account.
The hopping matrix elements 
\begin{equation}
       t_{ {\bf r}, {\bf r}+{\bf \Delta} }
       = \left\{ 
           \begin{array}{cc}
                         e^{\mp 2 \pi i  \alpha z}, &~~~ 
                         {\bf \Delta} \in \{ \pm {\bf e}_{y} \}, \\ 
                   
                         1,    &~~~
               {\bf \Delta} \in \{ \pm {\bf e}_{x}, \pm  {\bf e}_{z} \},  
           \end{array}
          \right. 
\end{equation}
describe a system with a homogeneous magnetic field $B$ in the 
$x$ direction and the Peierls phase $ \alpha = e B / h c $
is the number of flux quanta $ \Phi_{0} = hc/e $ per unit cell \cite{peie33}.
Here, the Landau gauge with the vector potential $ {\bf A} = (0,-Bz,0) $
is chosen.
The Hamiltonian (\ref{TBH}) is symmetric and periodic with 
respect to $ \alpha $ with a period $ \alpha=1 $ ($\Phi_{0}$ periodicity).
For a quasi-1D, bar-shaped system of cross section $ M \times M $ the  
Schr\"odinger equation $ H |\Psi \rangle = E |\Psi \rangle $
with  $ |\Psi \rangle = \sum  a_{{\bf r}} |{\bf r} \rangle $
can be written as a transfer matrix equation \cite{Pich81} 
$ {\bf u}_{z} =  {\bf T}_{z} {\bf u}_{z-1}$, where the vector
\begin{equation}
     {\bf u}_{z}=(a_{11z+1},\dots,a_{MMz+1};a_{11z},\dots,a_{MMz})^{T}
\end{equation}
contains the coefficients of the planes $ z $ and $ z+1 $  and 
\begin{equation}
{\bf T}_{z} = \left( \begin{array}{cc} 
                             {\bf Z}_{z} & -{\bf I} \\
                             {\bf I}     &   {\bf 0} 
                     \end{array}
               \right)
\end{equation}
is the $ 2M^{2} \times 2M^{2} $ transfer matrix. 
Each matrix $ {\bf Z}_{z}$ is of the order $ M^{2} \times M^{2}$ 
and contains the elements
of the Hamiltonian in the $ z $ plane as presented in 
Ref. \cite{Henn94}. The matrices $ {\bf I} $ connect successive 
$xy$ planes. The matrix $ {\bf Z}_{z} $ is a function of the energy 
$ E ,$ the magnetic field parameter $ \alpha$, and the random site energies 
$ \epsilon_{{\bf r}}$ on the slice $ z $.
 
In disordered quasi-1D systems all eigenstates are 
localized \cite{Pich81}. 
The wave function of a localized state decays exponentially with a 
localization length. The largest localization length $ \lambda_{M} $ can be 
identified as $ \gamma^{-1} $, the inverse 
of the smallest positive Lyapunov exponent (LE) 
$ \gamma = \min\{ \chi({\bf u}_{0}), \chi \ge 0 \} $ defined by \cite{osel68}
\begin{equation}
       \chi({\bf u} _{0}) = \lim_{L \to \infty }\frac{1}{L} 
       \ln \| {\bf T}_{L} \cdot \dots \cdot {\bf T }_{1} {\bf u}_{0} \|, 
\end{equation} 
where $ {\bf u}_{0} $ is the initial vector for $ z=0 $. 

 By using the Benettin algorithm \cite{osel68} we extracted 
$ \gamma $ from the approximate spectrum of LEs \cite{Pich81}.  
The localization length $ \lambda_{M} $ is a function of the 
cross section $ M $ of the bar, the electronic energy $ E, $  the number of 
flux quanta per unit cell $ \alpha$, and the disorder $ W $.

The critical behavior of $ \lambda_{M} $ near the MIT 
can be determined numerically by establishing the one-parameter scaling 
law \cite{Mack81}
\begin{equation} \label{scal}
     \frac{\lambda_{M}(E, W, \alpha)}{M} 
= f \left( \frac{ \lambda_{\infty} (E, W, \alpha) }{M} \right),
\end{equation} 
where $ \lambda_{\infty}  = \lim_{M \to \infty} \lambda_{M} $ 
is the localization length of the 3D system in the thermodynamic limit. 
The phase diagram of localization for a given $ \alpha $ 
describes the MIT in the $ (E,W)$ plane.
The mobility-edge trajectory $ E_{c}(W,\alpha) $ is determined by using 
the property that at the critical point the reduced localization 
length  of a quasi-1D system $ \lambda_{M}/M $ is independent of the cross 
section $M$, 
\begin{equation} \label{critpoint}
    \frac{\lambda_{M}(E_{c}, W, \alpha)}{M}  = {\rm const.} 
\end{equation}

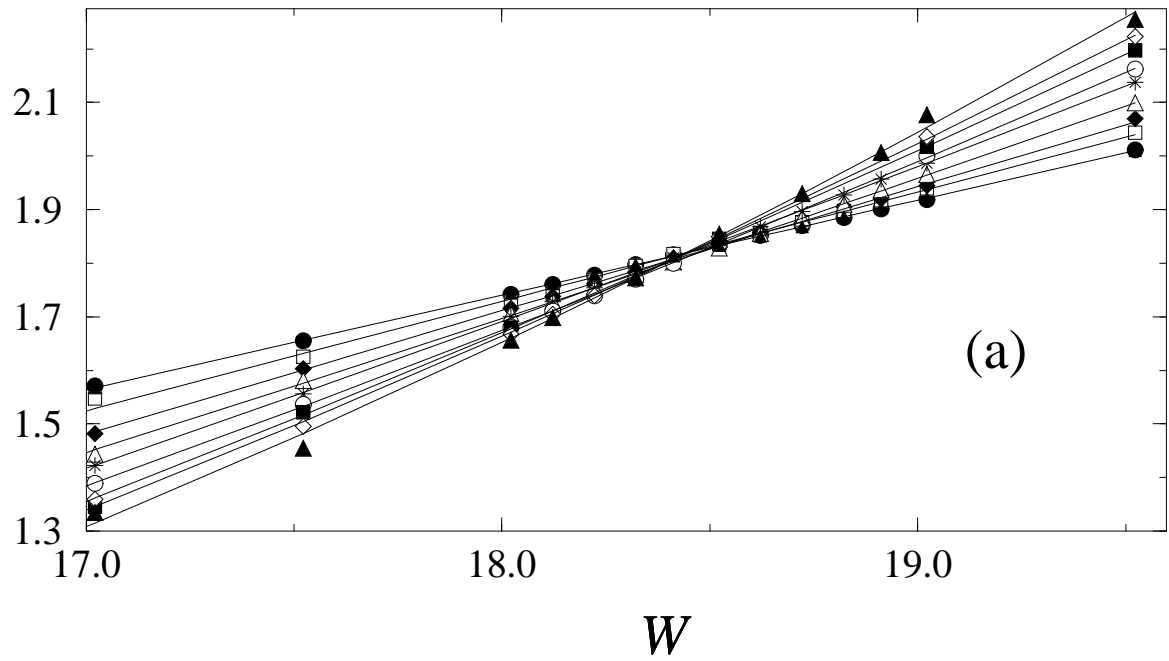
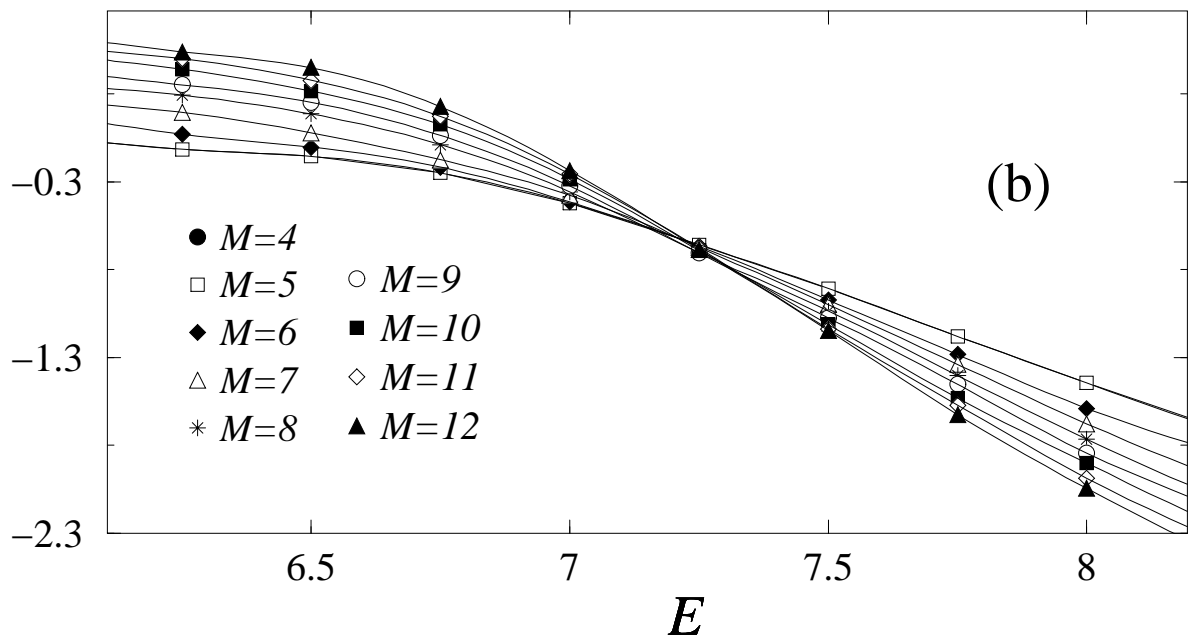
\begin{figure}[h]     
  \caption{ \label{Mitte2}
           Disorder and energy dependence of the reduced localization length 
           $ \lambda_{M}/M $ of quasi-1D disordered systems for various cross
           sections $ M $ in a magnetic field $ \alpha = 0.25$.
         a)~$ M/ \lambda_{M}$ vs. disorder W at the band center $ E=0$.
         b)~$ \ln(\lambda_{M}/M) $ vs. energy $ E $ at $W=12$ outside the
           unperturbed band. Lines are polynomial interpolations.
          } 
\end{figure}

 As  an example, Fig.\ref{Mitte2}(a) shows the inverse of the reduced 
localization length $ M/\lambda_{M}$ 
as a function of the disorder $ W$, at the band center $ E = 0.$ The 
Peierls phase was chosen to be $ \alpha= 0.25,$ at which 
the magnetic field has the strongest effect and the system belongs to 
the unitary universality class.
The data were calculated for different $ M $ 
with a statistical accuracy of $ 0.25 \%$. 
The MIT is indicated by the common crossing point at 
$ W_{c}(E=0,\alpha=0.25) = 18.35 \pm 0.11$. The sign  
of the size effect on $ \lambda_{M} $ changes when increasing the disorder 
from the metallic $( W < W_{c} )$ to the insulating regime $ ( W > W_{c})$. 
In the same way we locate the critical disorders for energies up to 
$ E=7$. By expanding Eq. (\ref{scal}) around the critical point at a given
energy and fitting the numerical data to the linearized form
\begin{equation}
\frac{ \lambda_{M} }{M} = \Lambda_{c} + A(W-W_{c})M^{1/\nu}, 
\end{equation}
one can extract the critical exponent $ \nu $ and the disorder dependence
of the localization length $ \lambda_{\infty}$, as has been performed  in 
Refs. \cite{Henn94} and \cite{Mack81}.
We have found $ \Lambda_{c} = 0.568  \pm 0.076 $ for $ \alpha=0 $ and 
$ \Lambda_{c} = 0.564  \pm 0.027 $ for $\alpha=0.25$ consistent with
data from Ref.~\cite{Slev97}.

 For energies beyond the unperturbed band $|E| > 6$, it is 
convenient to determine the critical points from the energy dependence 
of the localization length $ \lambda_{M} $ at fixed $ W$. 
In Fig.\ref{Mitte2}(b), the localization length is shown as a
function of the energy for various $ M$. The 
statistical accuracy is $ 1 \% $. The intersect of the lines 
signalizes critical behavior and yields the position of the mobility
edge. At $ W=12$, for example, $ E_{c}(\alpha=0.25) = 7.22 \pm 0.14 $.
 Other critical points $ E_{c}(W, \alpha) $ 
were determined similarly.

Combining all the data obtained by the finite-size scaling analysis 
as described above, one can construct the entire phase diagram of the
AM in the parameter space spanned by $ E $ and $ W $. 
Figure \ref{Trajektorie} shows the mobility-edge trajectory with and without 
a magnetic field. For zero field the states in 
region I are extended  whereas the states in II  are localized. 
The regimes I (metallic) and II (insulating) are separated by the 
mobility-edge trajectory (solid line).
The latter is modified when a magnetic field is applied. For 
$ \alpha = 0.25 $ states in region III become extended and the 
phase boundary shifts to higher values of critical disorder.
It is known from the theory of weak localization \cite{lee85} 
that in the presence of a magnetic field coherent 
time-reversed paths are eliminated and hence backscattering is suppressed.
Without a magnetic field, states in III are
localized by quantum interference effects. In this case
a weak magnetic field leads to a delocalization \cite{shap84,khme81}.
As a consequence, a stronger disorder is required in order to localize these
states (see Fig. 2). 
Thus, insulating systems with $ W $ slightly larger than the critical
disorder at zero field $ \left[ W > W_{c}(\alpha = 0) \right] $ undergo a 
transition to
a metal when the magnetic field is applied. This behavior is consistent with 
the mechanism of the field-induced MIT proposed by Shapiro \cite{shap84}. 

 The increase of the critical disorder $ W_{c} $ with the magnetic field 
$ \alpha $ at $ E= 0 $ is shown in the inset of Fig. \ref{Trajektorie}. 
For small $ \alpha $ the field dependence 
can be described by the following relation:
\begin{equation}
W_{c}(\alpha) - W_{c}(0) = 
(E_{c}(\alpha)-E_{c}(0)) \left. \frac{dW}{dE} \right|_{E=E_{c}(0)} 
\propto \alpha^{1/2 \nu}, 
\end{equation}
which was obtained previously using the scaling approach \cite{khme81}.
Our data for $ W_{c}(\alpha) $ are in agreement with this power law where
the critical exponent $ \nu \approx 1.4 $ is taken from 
Refs.~\cite{Slev97} and \cite{Henn94}. 
 
 In region IV, the trajectory moves into the metallic 
phase in contrast to region III.
Here, states that are extended for $ \alpha = 0 $ become localized for 
$ \alpha = 0.25$. This cannot be explained by the 
interference mechanism discussed above.
A qualitative understanding can be achieved by considering the 
motion of the trajectory of slightly disordered systems when applying a
magnetic field. In the limit $ W \to 0 $ the mobility edge merges with the 
band edge of an ordered system.  The zero-field value of the 
band edge is $ E_{b}(\alpha=0) = 6$. 
In a magnetic field, the band edge as a 
function of $ \alpha $ can be calculated 
numerically by using Hofstadter's algorithm \cite{hof76} generalized to 3D. 
The band shrinks if a magnetic field is applied and varies with a period 
of  one  flux quantum per unit cell, $ \alpha = 1 $ 
($\Phi_{0}$ periodicity). 
For example, for $ \alpha=0.25 $ the unperturbed band edge shifts to 
$ E_{b}(\alpha=0.25) \approx 4.8$ (Fig. \ref{Trajektorie}).
In slightly disordered systems this behavior 
persists. The band edge of such a system moves to a
lower value $ E_{b}(W,\alpha) <  E_{b}(W,0)$, indicating that the DOS for 
energies near the zero-field mobility edge 
decreases dramatically with increasing $\alpha$. Below a certain value
of the DOS an MIT is induced \cite{bulk85}.  In region IV,  
this causes a shift of the mobility 
edge to lower energies $ E_{c}(W,\alpha) <  E_{c}(W,0) $ 
when applying a magnetic field (Fig. \ref{Trajektorie}).

 In fact, the shift of the phase boundary is due to the combination 
of both types of field effect mentioned above which compete with each other.
Close to the band center (III), where the DOS changes negligibly with $ E$, 
the interference effect dominates. On the other hand, in region
IV, close to the band tails, 
the DOS effect is much stronger than the interference effect. Here, the
behavior of $ E_{c}(\alpha) $ is determined mainly by the
energy and field dependence of the DOS. For example, for $ \alpha = 0.25 $
we find an intersect of the two phase trajectories at $ W^{*} \approx 14.5 $ 
and $ E^{*} \approx 7.8$, where the two effects are of the same order of 
magnitude. We believe that a similar field dependence of the phase 
diagram is also valid for a Gaussian distribution of on-site energies 
$ \epsilon_{{\bf r}}$.

 We now again concentrate on the band center $E=0$ in order to investigate 
the field dependence of the critical parameters.
Figure \ref{peipha} shows the  localization length
$ \lambda_{M}  $ of quasi-1D systems 
as a function of magnetic field for various $ W $ corresponding 
to the delocalized, critical, and localized regime in zero field, respectively.
The data were calculated with a precision of $ 0.25 \%$. 
For $ \alpha $ different from half integers, $ \lambda_{M}(E=0)$ 
is larger than without a magnetic field.
This field-induced enhancement of the localization length is directly 
related to the shift of the phase boundary in region III.
One sees in Fig. 3 that for $ E=0, W \ge 16.5$, and $ M=4 $ the localization 
lengths $ \lambda_{M} $ coincide  for both $\alpha = 0$ and $ \alpha=0.5 $ 
within the statistical uncertainties.  
They   vary with a period of {\it half a flux 
quantum} per unit cell ($\Phi_{0}/2$ periodicity).
This was also checked for other widths of $ M=5, \dots, 12 $ 
by comparing $ \lambda_{M}(E=0,\alpha = 0) $ with  
$ \lambda_{M}(E=0,\alpha = 0.5) $.
In the insulating region our results are consistent with a periodically 
varying transition amplitude of strongly localized electrons as derived 
analytically with a directed path method by Lin and Nori \cite{Lin96}. 
However, we find that the $\Phi_{0}/2$ periodicity does not persist for 
$ E \not= 0$ as shown in Fig. 3. Thus the former 
seems to be an intrinsic property of the band center, 
%
%
around which
the on-site energies $\epsilon_{ {\bf r} } $ are distributed 
symmetrically~\cite{Carini84}. 
Furthermore, for the investigated systems of width $ M=4, \dots , 12$
we observe 
that $\lambda_{M}(\alpha=0.5) < \lambda_{M}(\alpha=0)$, though 
for both values of $ \alpha $ the Hamiltonian belongs to the orthogonal 
universality class. Assuming that $ \Lambda_{c} $ does not change in this case,
one obtains that for $ \alpha=0.5 $ and $ E \not= 0 $ the MIT should 
unexpectedly  occur at a lower disorder than in the zero field case, 
$ W_{c}(\alpha=0.5) < W_{c}(\alpha=0)$.

The maximum-entropy ansatz predicts the increase of the
localization length in quasi-1D systems by a universal factor of $ 2$, when 
breaking the time reversal symmetry \cite{Ston91}. 
Since in the present calculations the parameters are beyond the range of 
validity of this universal relation, it would be desirable to extend the 
investigations to this regime for checking the prediction. Another 
interesting problem is to study the disappearance 
of this relation, as was argued in Ref.~\cite{Lern95}, for $ d \ge 2$,
when extrapolating to 3D systems by scaling the
cross section $ M $ of the quasi-1D bar.

 In conclusion, we have numerically calculated the phase diagram of 
localization in 3D disordered systems in the presence of a magnetic field. 
Comparing the obtained diagram with the zero-field result, we identify two 
regimes with different magnetic field dependence of the phase boundary.
In the band center, 
the phase boundary is shifted towards higher values of critical 
disorder, so that the metallic phase is broadened. This is mainly due to the 
suppression 
of the interference of time-reversed paths by a magnetic field, leading to the 
delocalization of electron states. On the other hand, close to the band tails 
the location of the MIT in slightly disordered systems $(W \to 0)$
is dominated by the field dependence of the DOS. Here, the mobility edge 
shifts towards smaller energies, thus {\em diminishing} the metallic phase.
Our numerical findings for small fields 
show that the behavior of the critical disorder
is consistent with predictions by scaling theory \cite{khme81}.
Thus, we have shown that the phase diagram of localization is  
influenced by an external perturbation which breaks the time reversal 
invariance.
                                         
 Discussions with D. Belitz are gratefully acknowledged.
This work was supported by DFG-Projekt No. Kr627/8-1, the Graduiertenkolleg
``Physik nanostrukturierter Festk\"orper'', University of Hamburg, and 
NATO Grant No. CRG 941250.

\newpage

\begin{figure}[h]     
  \caption[]{ \label{Trajektorie}
           The phase diagram of the Anderson model of localization. 
           Full dots show the critical points $ \{ E_c, W_c \} $ for 
           a magnetic field $\alpha=0.25$.
           The error bars indicate the statistical uncertainties
           due to the finite length of the quasi--1D systems.
           The open circles show numerical results for $ \alpha= 0$ 
           taken from Ref.~\cite{bulk85}.   
           The full and the dotted line represent the mobility edge trajectory 
           for $ \alpha=0$ and $ \alpha=0.25 $ respectively. 
     Inset:
           critical disorder $ W_{c} $ vs. magnetic field $ \alpha $ in the
           band center $ E=0$. The curve shows the power-law behavior from 
           scaling arguments~\cite{khme81}.}  
\end{figure}

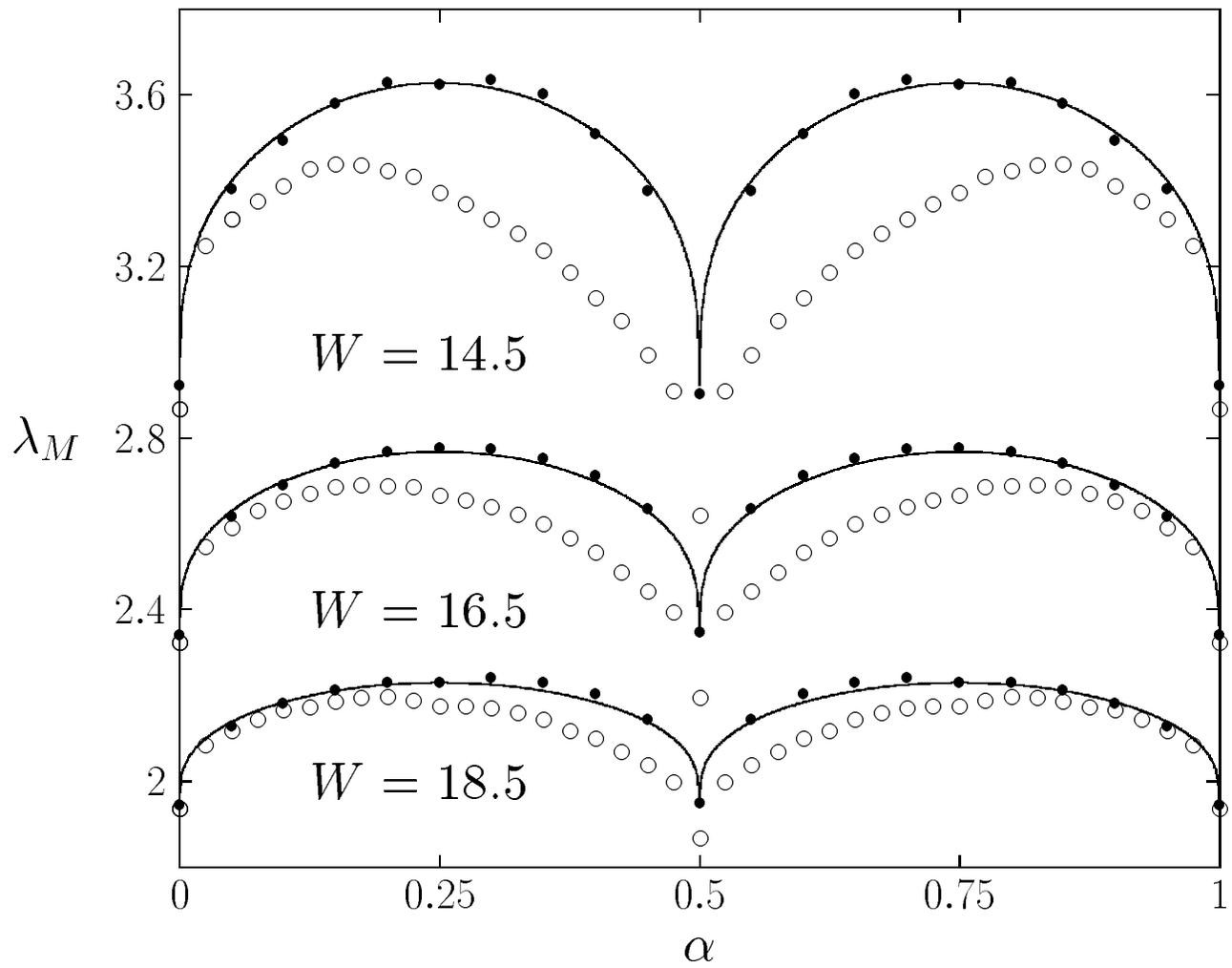
\begin{figure}[h]     
  \caption{ \label{peipha} 
       Localization length $ \lambda_{M} $ of a quasi-1D disordered system
       with a cross section $ M =4 $  
       for various disorders $ W $ corresponding to different regimes 
       (extended states, critical region and localized states)
       as a function of the number of flux quanta per unit cell $ \alpha $ for 
       $ E=0 $ (full dots) and $ E=4 $ (open circles). 
       The continuous lines are fits by $ \Phi_{0}/2$-periodic functions.
       }
\end{figure}             

\end{document}